\documentclass[prb,aps,amsmath,twocolumn,superscriptaddress,a4paper]{revtex4}
\pdfoutput=1
\usepackage{hyperref}

\newcommand{\vek}[1]{{\vec{\mathbf{#1}}}}     
\def \d{\mathop{{\rm d}\!}\nolimits}
\def \der #1#2{\frac{\d #1}{\d #2}}
\def \derv #1#2#3{\left.\der{#1}{#2}\right|_{#3}}

\def \parder #1#2{\frac{\partial #1}{\partial #2}}

\def \thh {\ensuremath{{\theta^*}}}
\newcommand{\dpiint}{\frac{1}{2\pi{}}\int}
\newcommand{\os}{\ensuremath{\overline{s}}}
\newcommand{\bpertn}{{\tilde{b}^1_n}}
\newcommand{\bpertngeo}{{\tilde{b}^{\prime 1}_n}}
\newcommand{\bpert}[2]{{\tilde{b}^1_{({#1},{#2})}}}
\newcommand{\bpertreal}[2]{{b^1_{({#1},{#2})}}}
\newcommand{\Hpert}[2]{{\tilde{H}_{({#1},{#2})}}}
\newcommand{\Hpertreal}[2]{{{H}_{({#1},{#2})}}}
\newcommand{\bpertgeo}[2]{{\tilde{b}^{\prime 1}_{({#1},{#2})}}}
\newcommand{\mgeo}{{\overline{m}}}
\newcommand{\psinorm}{{\psi_{N}}}
\newcommand{\toroflux}{{\Phi}}
\newcommand{\conjg}[1]{{{#1}^{*}}}
\newcommand{\homoc}{\phi{}}
\def \enorm #1{\ensuremath{{\hat{\vek{e}}_{#1}}}} 
\begin{document}
\title{Resonant magnetic perturbations and divertor footprints in poloidally diverted tokamaks}

\author{Pavel Cahyna}
 \email{cahyna@ipp.cas.cz}
 \affiliation{Institute of Plasma Physics AS CR v.v.i., Association EURATOM/IPP.CR, Prague, Czech Republic}
\author{Eric Nardon}%
 \affiliation{Association EURATOM-CEA, CEA/DSM/IRFM, CEA Cadarache, 13108 St-Paul-lez-Durance, France}
\begin{abstract}
  General formula describing both the divertor strike point
  splitting and width of magnetic islands created by resonant magnetic
  perturbations (RMPs) in a poloidally diverted tokamak equilibrium is
  derived. Under the assumption that the RMP is produced by coils at
  the low-field side such as those used to control edge localized
  modes (ELMs) it is demonstrated that the width of islands on
  different magnetic surfaces at the edge and the amount of divertor
  splitting are related to each other. Explanation is provided of
  aligned maxima of the perturbation spectra with the safety factor
  profile -- an effect empirically observed in models of many
  perturbation coil designs.
\end{abstract}

\maketitle

\section{Introduction}
Resonant magnetic perturbations (RMPs) are being investigated as tool
to control edge localized modes (ELMs), in particular their
application to ITER is foreseen. The RMPs for ELM control are produced
by coils whose design is specific to each tokamak. This method is
reminiscent of the ergodic divertor on tokamaks Tore Supra and TEXTOR
which also essentialy relies on RMPs produced by dedicated systems of
coils, and many aspects are similar, especially the formation of
magnetic islands on rational surfaces and possible stochastic
transport when the islands overlap. However the presence of X-point
in the poloidally diverted tokamaks provides some unique features: the
splitting of divertor strike points, or divertor footprints -- a
signature of the homoclinic tangle created by
the perturbation\cite{Evans05}, and the divergence of the safety
factor profile at the separatrix, due to which the number of rational
surfaces even for one toroidal mode is infinite and island overlap on
them is facilitated. The noncircular geometry of the plasma
cross-section also complicates analytical treatment of
magnetic islands, requiring cautious use of non-orthogonal coordinate
systems.

Since both the homoclinic tangle and magnetic islands with the
resulting stochasticity are consequences of the perturbation, it is
natural to ask if there is any relation between them. It has been
known that while every island chain is related to a different polodal
mode of the perturbation, their sizes can be expressed by a single
function -- the Poincar\'e-type integral which also generalizes to an
expression for the length of the divertor
footprints\cite{abdullaev:042508}.The present paper uses instead the
more familiar formalism of the Melnikov integral\cite{wiggins} (a
standard tool for the analysis of the homoclinic tangle) and explores
the radial dependence of this function in the often encountered case
of RMPs from the perturbation coils localized at the low-field side
(LFS), as it is the case for most ELM control coil designs.

The paper starts with a review of the method to calculate island
sizes. Care is taken to provide a formula valid in a plasma with a
general aspect ratio and non-circular cross-section as this is
essential for the edge pedestal region near the separatrix which is
crucial for ELM control. It is shown how the island sizes are
determined by a function whose definition is not affected by the
coordinate singularity on the separatrix, and which we call the
Melnikov-like function. The homoclinic tangle and divertor footprints
are then explained together with the method of Melnikov integral for
determining analytically the divertor footprints length. A
particularily simple expression is given for the case when the
perurbation has only one dominant toroidal mode. The relation between
Melnikov integral and the Melnikov-like function is explained. Then we
restrict our treatment to the case of perturbation localized on the
LFS. Under this assumption the relation between modes at the different
surfaces is derived and it is shown how does the divertor footprint
length relate to the sizes of magnetic islands at the edge.

\section{Width of magnetic islands in a realistic geometry}
\subsection{Expression using the resonant modes of the perturbation}
Nonaxisymmetric magnetic perturbations of a tokamak magnetic field
produce chains of magnetic islands on magnetic surfaces with low-order
rational values of the safety factor $q$. Those islands are created by
resonances of the perturbation field $\delta \vek{B}$ with the
unperturbed field lines on the rational surface. To express these
resonances we use a magnetic coordinate system \cite{park:064501} $(s,
\thh, \varphi)$ where $s$ is a flux surface label, $\thh$ the poloidal
coordinate and $\varphi$ the toroidal coordinate. The angular
coordinates $\thh, \varphi$ are chosen as in the PEST coordinates
\cite{PEST} and the radial coordinate $s$ is defined as square root of
the normalized poloidal magnetic flux $\psinorm$: $s=\sqrt{\psinorm}$, in
accordance with previous works
(e.g. Refs.~\onlinecite{marinanf2005,marinanf2008}). The coordinate $\varphi$ is
simply the geometric toroidal angle and the coordinate system has the
property that $\thh=o\varphi/q(s)+\mbox{const.}$ along a field
line, where $o$ represents the orientation of the magnetic field:
$o=1$ for the left-handed field and $o=-1$ for the right-handed
field\footnote{The orientation of
  the magnetic field depends on the relative orientation of plasma
  current and toroidal field: parallel in a right-handed field case,
  antiparallel in a left-handed field case.}.
The islands are created by the $(m, n)$ Fourier component 
$\bpert{m}{n}$ of
the normalized perturbation $\delta b^1\equiv{B^1}/{B^3}$, where
$B^1\equiv\delta \vek{B}\cdot\nabla s$ is the contravariant $s$
component of the perturbation and $B^3\equiv\vek{B}\cdot\nabla \varphi$ is
the contravariant $\varphi$ component of the equilibrium field. The
Fourier transform is taken with respect to the $\thh$ and $\varphi$
coordinates, thus we have
\begin{equation}
\label{eq:bmnsum}
\delta b^1=\sum_{m,n=-\infty}^{\infty}{\bpert{m}{n}\exp{[\mbox{i}(m\thh-n\varphi)]}}
\end{equation}
and the Fourier harmonics can be obtained as
\begin{equation}
  \label{eq:mnmodes}
  \bpert{m}{n} = \frac{1}{(2\pi)^2}\oint
  \exp[-\mbox{i}(m\thh - n\varphi)] \delta b^1\d\thh\d\varphi.
\end{equation}
A Fourier component is resonant with the unperturbed field lines
when $q=om/n$.
The values $\bpert{m}{n}$ are complex, and as $\delta b^1$ is real, the
following relation for the complex conjugate holds:
$\conjg{\left(\bpert{m}{n}\right)} = \bpert{-m}{-n}$. An alternative to
(\ref{eq:bmnsum}) is a representation using purely real coefficients:
\begin{eqnarray}
  \label{eq:realbmnsum}
  \delta b^1&=&\sum_{m=-\infty,n=1}^{\infty}
  {\bpertreal{m}{n}\sin(m\thh-n\varphi+\chi_{mn})}\\
  \bpertreal{m}{n} &=& 2|\bpert{m}{n}|\\
  \chi_{mn} &=& \arg\bpert{m}{n}
\end{eqnarray}

Widely used formulae exist for determining the width of magnetic islands
from the Fourier spectrum of the perturbation
\cite{wesson,marinanf2005}. They are typically derived in a
cylindrical geometry where the toroidal curvature is not being taken
into account (the toroidal magnetic field is considered constant),
thus $B^3$ in the expression for $\delta b^1$ is approximated by its
value at the magnetic axis. As noted in Ref~\onlinecite{Kaleck1999}, this
leads to an error in estimating the island width. In the example of
TEXTOR and its dynamic ergodic divertor (DED), the island size was
overestimated because the DED coils are located at the high-field
side, thus the actual value of $B^3$ is larger and $\delta b^1$ is
smaller than in the cylindrical approximation. For the ergodic
divertor of Tore~Supra, which was located at the low-field side, the
island sizes were underestimated. It should be noted that while the
toroidal field magnitude $B_{\text{T}}$ varies with the radial
distance $R$ from the major axis as $B_{\text{T}} \propto 1/R$, for
the contravariant component the dependence is stronger: $B^3 \propto
1/R^2$.

Moreover the cylindrical formula for island widths uses $r$ (the
distance from the magnetic axis) as a radial coordinate and thus is
valid only in a situation where the magnetic surfaces have circular
and concentric cross-sections. In divertor tokamaks we are far from
this geometry, especially in the edge region near the separatrix which
is the most important when perturbations are used as an ELM control
mechanism. We thus need a formula which would be usable in a general
geometry, with a varying toroidal field and noncircular flux surfaces,
using for example the coordinate $s$ as a general flux surface label
instead of $r$. 

To derive this formula we introduce new coordinates
$\chi\equiv\thh-n/m\varphi$ and $\os\equiv s-s_0$ where $s_0$ is the
flux label of the resonant surface where $q=m/n$. The differential
equation of the field line are
\begin{eqnarray}
  \label{eq:sline}
  \der{s}{\varphi} &=& \frac{B^1}{B^3} \\
  \label{eq:thline}
  \der{\thh}{\varphi} &=& \frac{1}{q}
\end{eqnarray}
Using the coordinates $\chi$ and $\os$ and the relation
\begin{equation}
  \label{eq:dchidphi}
  \der{\chi}{\varphi} = \frac{1}{q} - \frac{n}{m}
\end{equation}
the equation (\ref{eq:sline}) becomes
\begin{equation}
\label{lineeq}
1/q - n/m\,\d\os = {B^1}/{B^3}\,\d\chi
\end{equation}
Keeping only the resonant part of the perturbation, thus substituting
${B^1}/{B^3}$ by $\bpertreal{m}{n} \sin(m\chi+\chi_{mn})$,
we obtain
\begin{equation}
\label{reslineeq}
1/q - n/m\,\d\os = \bpertreal{m}{n} \sin(m\chi+\chi_{mn})\,\d\chi
\end{equation}
Using a linear approximation of the left side, we obtain
\begin{equation}
\derv{q^{-1}}{\os}{q=m/n}\os\,\d\os \approx \bpertreal{m}{n} \sin(m\chi+\chi_{mn})\,\d\chi\label{perturbeq}
\end{equation}
This equation can be easily integrated to obtain an algebraic equation
for field lines:
\begin{equation}
  \label{eq:algline}
  \os^2 \approx \frac{2q^2\bpertreal{m}{n}}{q^\prime m} [\cos(m\chi+\chi_{mn}) + C]
\end{equation}
where $q^\prime \equiv \d q/\d s$ at the resonant surface and $C$ is
an integration constant. The choice $C=1$ corresponds to the island
separatrix whose maximum radial excursion is the island half-width $\delta$,
given by the formula
\begin{equation}
  \label{eq:linhalfwidth}
  \delta = \sqrt{\frac{4q^2\bpertreal{m}{n}}{q^\prime m}}
\end{equation}

An alternative to this approach is to use a Hamiltonian approach where
the field lines are interpreted as trajectories of a Hamiltonian
dynamical system whose Hamilton function is the poloidal flux and the
perturbation is represented as a perturbed Hamiltonian (flux). This
approach has been used in many theoretical works. We briefly review it
in the appendix~\ref{sec:hamilt-repr-field} and prove its equivalence
to the approach described above. It should be emphasized that the
hamiltonian approach automatically includes correctly the effects of
toroidal geometry and non-circular cross-section -- no corrections are
necessary. It is however still important to have a correct formula
using the perturbed magnetic field (\ref{eq:linhalfwidth}), because
this is the approach usually used in numerical studies of perturbation
coil designs, as the perturbed field can be readily calculated from
the coil geometry by the Biot-Savart formula. We will also see that
the harmonics of the perturbed field are directly related to the
Melnikov function.

\subsection{Expression using Melnikov-type integral}
\label{sec:expr-meln-int}

The coordinate system $(\thh, s)$ on a poloidal plane has a
singularity at the separatrix. It is useful to define a value
characterizing magnetic islands which, unlike $\bpert{m}{n}$, will not
use the $\thh$ coordinate, so it will stay well-defined even at the
separatrix.

We start by defining a coordinate $\homoc$ which will be used instead
of $\thh$. We follow a procedure used in the definition of the
separatrix map and the Melnikov
integral\cite{zaslavsky2005,wiggins}. For every magnetic surface the
point on the outboard midplane has $\homoc = 0$. Following a field
line parameterized by the toroidal angle $\varphi$ from this point, we
assign to any other point on the field line the value $\homoc =
\varphi$. Thus $\homoc$ of a given point is the toroidal angle needed
to reach it by following a field line from the outboard
midplane. Since the field line returns to the same poloidal position
after making $q$ toroidal turns, the range of the coordinate $\homoc$
needed to cover a magnetic surface in the poloidal plane is $(-q
\pi,q\pi)$ where the endpoints of this interval are identified with
each other. Together with a flux surface label such as $s$ we obtain a
coordinate system on the poloidal plane. The separatrix is a special
case: it is covered by $\homoc \in (-\infty, \infty)$ since $q$ is
infinite on the separatrix, and the X-point corresponds to $\homoc =
\pm\infty$. In this case $\homoc$ is called a \emph{homoclinic
  coordinate}\cite{wiggins}. As on a field line $\varphi = oq\thh +
\text{const.}$ and $\thh = 0$ on the outboard midplane where
$\homoc=0$, the relation between $\thh$ and $\homoc$ is $\homoc =
oq\thh$. Using this relation, the definition (\ref{eq:mnmodes}) of
$\bpert{m}{n}$ can be rewritten as
\begin{multline}
\label{eq:homocmodes}
  \bpert{m}{n} = \\
\frac{1}{oq(2\pi)^2}\int_{-q\pi}^{q\pi}\int_0^{2\pi}
  \exp\left[-\mbox{i}\left(o\frac{m}{q}\homoc - n\varphi\right)\right]
  \delta b^1\d\varphi\d\homoc
\end{multline}
If we define a toroidal perturbation Fourier mode $\bpertn$ as
\begin{equation}
\label{eq:bpertn}
  \bpertn(\homoc) = \frac{1}{2\pi}\int_0^{2\pi}
  \exp(\mbox{i}n\varphi)
  \delta b^1(\homoc, \varphi)\d\varphi
\end{equation}
we may write (\ref{eq:homocmodes}) as
\begin{equation}
\label{eq:homocnmode}
  \bpert{m}{n} = \frac{1}{o2\pi q}\int_{-q\pi}^{q\pi}
  \exp\left[-\mbox{i}o\left(\frac{m}{q}\homoc\right)\right]
   \bpertn(\homoc)\d\homoc
\end{equation}
On resonant surfaces with $q=om/n$ this may be simplified to
\begin{equation}
\label{eq:homocresmode}
  \bpert{m=onq}{n} = \frac{1}{o2\pi q}\int_{-q\pi}^{q\pi}
  \exp(-\mbox{i}n\homoc)
  \bpertn(\homoc)\d\homoc
\end{equation}
($\bpertn$, $\bpert{m}{n}$ and  $\delta b^1$ all depend also on the
magnetic surface, this was omitted from the expressions above for
brevity). In (\ref{eq:homocresmode}) there appears a complex
Melnikov-like function $\tilde{S}_n(s)$ given by
\begin{equation}
  \label{eq:melnikov-like}
   \tilde{S}_n(s) \equiv 
\int_{-q\pi}^{q\pi} \exp(-\mbox{i}n\homoc)\bpertn(s, \homoc)\d\homoc,
\end{equation}
defined using the coordinate $\homoc$ which does not have a
singularity at the separatrix, so the definition can be extended to
the separatrix:
\begin{equation}
\label{eq:Mnsepar}
  \tilde{S}_n(s=1) \equiv \int_{-\infty}^{\infty}
  \exp(-\mbox{i}n\homoc)
  \bpertn(s=1, \homoc)\d\homoc.
\end{equation}
The function $\tilde{S}_n$ fulfills our requirement: it can
replace $\bpert{m}{n}=\tilde{S}_n/(o2\pi q)$ and is defined using
values which remain
regular at the separatrix. The island width (\ref{eq:linhalfwidth})
can be expressed using $\tilde{S}_n$ instead of $\bpert{m}{n}$:
\begin{equation}
\delta = \sqrt{\frac{4|\tilde{S}_n|}{n\pi q^\prime}}.\label{eq:linhalfwidth-Melnikov}
\end{equation}
The only remaining divergent term in (\ref{eq:linhalfwidth-Melnikov})
is the shear $q^\prime$ which grows to infinity at the separatrix.
This dependence is physical: its consequence is that island width has
a zero limit at the separatrix.

\section{Divertor footprints}
Since the particle and heat transport are mostly parallel to field
lines, the patterns of particle and heat flux to the divertor plates
can be expected to be related to the divertor magnetic footprints,
i.e. the patterns of intersections of field lines with the
divertor. Field lines which carry heat and particle fluxes from inside
the plasma are those with a high connection length, i.e. the number of
toroidal turns following the field line in the plasma before it
reaches the wall again.

Since the field lines can be interpreted as trajectories of a
Hamiltonian dynamical system with the toroidal angle in the role of
the time, methods of the theory of Hamiltonian systems can be used. A
concept especially useful for the study of divertor footprints is the
one of invariant manifolds\cite{Evans05}. An invariant manifold is a surface in
the phase space of the dynamical system which remains invariant by the
time evolution of the system, thus a trajectory with an initial point
on the invariant manifold is constrained to remain on it \cite{wiggins}. In our
case the trajectories are field lines and one example of invariant
manifolds are the magnetic surfaces of the toroidally symmetric
tokamak equilibrium. A particularly interesting case of invariant
manifolds are the stable and unstable manifolds of hyperbolic fixed
points. A stable manifold is formed by field lines asymptotically
approaching the fixed point, while the unstable manifold is formed by
field lines asymptotically leaving the fixed point. The definition
depends on the direction in which the field lines are followed. If we
follow them in the opposite direction, the stable manifold becomes
unstable and vice versa. In plasma equilibria the hyperbolic fixed
points are called X-points and are associated with the poloidal
divertor or with magnetic islands. An example of invariant manifolds
to a fixed point is the separatrix of a toroidally symmetric
configuration with a poloidal divertor. Here the stable and unstable
manifolds coincide to form the separatrix. When a perturbation
appears, the separatrix splits into the stable and unstable manifolds
which no longer coincide, but intersect transversally infinitely many
times. Close to the X-point in the direction from which the field
lines approach it (the stable direction) the stable manifold is close
to the unperturbed separatrix, but the unstable manifold wildly
oscillates, creating lobes that become longer and narrower when the
X-point is approached \cite{Evans05}. In the direction of field lines leaving the X-point (the unstable direction) we obtain a similar picture with the roles of the stable and unstable manifolds reversed. This complex structure is called a homoclinic or heteroclinic tangle.
An important property of the invariant manifolds is that field lines
can't cross them, because field lines can't intersect. Invariant
manifolds thus act as boundaries for the field lines. Field lines
originating in the hot plasma core are contained inside the invariant
manifolds of the X-point and the only way they can reach the divertor
targets is when the lobes of the (un)stable manifolds near the X-point
intersect the target plates \cite{WingenPoP09}. By tracing the
intersection of the manifolds with the plates one obtains curves which
delimit the region connected to the plasma core, characterized by
mostly high connection length in the laminar plot. Those divertor
footprints typically take the form of long spiralling bands, each band
corresponding to the intersection of a protruding lobe of a stable or
unstable manifold with the divertor. 

The divertor footprints have a complicated inner structure and not all
points inside the manifolds have high connection length. Some of them
are connected to the opposite divertor plate after two poloidal turns
by laminar flux tubes which do not penetrate deeply under the
separatrix. The points with high connection lengths are the images of invariant manifolds of the X-points of the inner island chains \cite{WingenPoP09}. 
This fine structure was studied in detail in \cite{WingenNF09}. Here we
focus on the on the overall shape of the divertor footprints which is
given by the invariant manifolds of the divertor X-point and is better experimentally accessible.

The length of the spiral can be characterized by the maximum value of $s$
reached, i.e. the value $s_\text{tip}$ at its tip. The difference
$\Delta{}s_\text{max}$
of $s_\text{tip}$ and the separatrix value $s = 1$
expresses the radial distance on the divertor plate between the
footprint's tip and the unperturbed strike point, which lies at the
intersection of the unperturbed separatrix with the divertor
plate. The unstable manifold\footnote{Or the stable manifold,
  depending on which strike point we consider -- the inner or the
  outer one.} is the footprint's boundary and so the footprint's tip
is the point on the manifold which is the most distant from the
unperturbed separatrix, the distance in terms of $s$ being
$\Delta{}s_\text{max}$. The value $\Delta{}s_\text{max}$ thus
quantifies the magnitude of the separatrix splitting.

To estimate the separation between the unperturbed separatrix and the
unstable manifold we will follow two field lines -- one in the
unperturbed field, lying on the separatrix, and the other in the
perturbed field, lying on the unstable manifold. They are
parameterized by the toroidal angle $\varphi$. Let us choose them so
that they are initially (in the vicinity of the X-point which they
approach asymptotically when followed backwards in $\varphi$) close to
each other. The parametric equations of the perturbed field line are
\begin{eqnarray}
  \label{eq:parampert}
  s=s^\prime(\varphi) \\
  \homoc = \homoc^\prime(\varphi).
\end{eqnarray}
For the unperturbed field line they can be written explicitely using
the definition of $\homoc$ and the fact that the unperturbed field
line lies on the separatrix where $s=1$:
\begin{eqnarray}
  \label{eq:paramunpert}
  s=s(\varphi) = 1 \\
  \homoc = \homoc(\varphi) = \varphi - \homoc_0
\end{eqnarray} 
where the constant $\homoc_0$ determines the toroidal phase: it is
the toroidal angle of the point where the field line crosses the
outboard midplane.

The rate of change of $s$ along the perturbed field line
is
\begin{equation}
  \label{eq:sder}
  \der{s^\prime}{\varphi} = \delta b^1(s^\prime(\varphi),
  \homoc^\prime(\varphi), \varphi)
\end{equation}
If the perturbed field line does not deviate significantly as a result
of the perturbation, we may use a first-order perturbative
approximation and evaluate $b^1$ on the unperturbed field line:
\begin{equation}
  \label{eq:sderapprox}
  \der{s^\prime}{\varphi} = \delta b^1(s = 1, \homoc(\varphi), \varphi)
\end{equation}

The deviation $\Delta s$ of the perturbed field line from the
unperturbed one after a full poloidal turn is given by
the integral of (\ref{eq:sderapprox}):
\begin{equation}
  \label{eq:deltas}
  \Delta s(\homoc_0) = \int_{-\infty}^\infty \delta b^1(s = 1, \homoc(\varphi) =
  \varphi - \homoc_0, \varphi) \d
  \varphi
\end{equation}
or using $\homoc$ as the parameter:
\begin{equation}
  \label{eq:deltas-homoc}
  \Delta s(\homoc_0) = S(\homoc_0) \equiv \int_{-\infty}^\infty \delta b^1(s = 1, \homoc,
  \homoc+\homoc_0) \d\homoc.
\end{equation}
Note that $\Delta s(\homoc_0)$ is a function of the toroidal phase
$\homoc_0$ and may be zero: this happens when the unstable manifold
intersects the unperturbed separatrix.

The function $S$ defined by the integral (\ref{eq:deltas-homoc}) is
closely related to the Melnikov function $M$:
\begin{equation}
  \label{eq:melnikov}
  M(\homoc_0) = \int_{-\infty}^\infty \delta b^{\psi}(s = 1, \homoc,
  \homoc+\homoc_0) \d\homoc = \der{\psi}{s} S(\homoc_0)
\end{equation}
where $\delta b^{\psi}(s = 1, \homoc, \homoc+\homoc_0) = \der{\psi}{s}
b^1(s = 1, \homoc, \homoc+\homoc_0)$ is the contravariant component of
$\delta B$ with respect to the $\psi$ coordinate. The only difference
between $M$ and $S$ is that $S$ gives the change of $s$ while $M$
gives the change of $\psi$\cite{wiggins}.

If there is only one toroidal mode $\bpertn(\homoc)$ of the
perturbation [cf. equation (\ref{eq:bpertn})], the function
$S$ can be replaced by a single complex number $\tilde{S}_n(s=1)$:
\begin{eqnarray}
  S(\homoc_0) &=&  \int_{-\infty}^\infty \delta b^1(s = 1, \homoc,
  \homoc+\homoc_0) \d\homoc \nonumber \\
  &=& \int_{-\infty}^\infty
  2\Re\left\{\exp[-\mbox{i}n(\homoc+\homoc_0)]\bpertn(s = 1, \homoc)\right\}
  \d\homoc \nonumber \\
  &=& 2\Re\left[\exp(-\mbox{i}n\homoc_0)
    \int_{-\infty}^\infty
    \exp(-\mbox{i}n\homoc)\bpertn(s = 1, \homoc)
    \d\homoc\right] \nonumber \\
  \label{eq:Scomplex}
  &=& 2\Re\left[\exp(-\mbox{i}n\homoc_0)\tilde{S}_n(s=1)\right]
\end{eqnarray}
$\tilde{S}_n(s=1)$ is defined by the equation (\ref{eq:Mnsepar}),
which also naturally extends the definition of $\tilde{S}_n(s)$ to the
domain $s<1$. Analogously the value
$\tilde{M}_n(s)\equiv\der{\psi}{s}\tilde{S}_n(s)$ can be used to
express the Melnikov function $M$ as a single number
$\tilde{M}_n(s=1)$.

The value $\Delta{}s_\text{max}$ is the maximum deviation of the
unstable manifold: $\Delta{}s_\text{max} =
\max_{\homoc_0}\Delta{}s(\homoc_0)$. Using $\tilde{S}_n(s=1)$ it is
expressed as
\begin{equation}
  \label{eq:deltasmax}
  \Delta{}s_\text{max} = 2 |\tilde{S}_n(s=1)|
\end{equation}
The island widths and the magnitude of separatrix splitting, and
consequently the length of the divertor footprints, are given by a
single function $\tilde{S}_n(s)$: the island widths by its values at
$s<1$ and the magnitude of splitting by the value at $s=1$.

It has been already known that island widths and the magnitude of
splitting can be described by a signle radial
function\cite{abdullaev:042508}: the Poincar\'e-type integral $R_n$
which is an integral of the modes of the perturbed poloidal flux
$H_1$ (cf. (\ref{eq:hammnsum}))
instead of the perturbed field $\bpertn$. It can be shown that
$\tilde{M}_n=\epsilon nR_n$ so for a single toroidal mode of the
perturbation our formalism of the Melnikov-like functions $S$ or $M$
is equivalent to the Poincar\'e-type integral approach. The
reference~\onlinecite{abdullaev:042508} gives also other results
expressed in terms of the function $R_n$ such as the width of the
stochastic layer and the field line diffusion coefficient which can be
also simply reformulated using Melnikov-like functions.

\section{Specific form of the modes of a localized low-field side perturbation}
In the previous sections we introduced the Melnikov-like function
$\tilde{S}_n$ and showed how it expresses both the sizes of the
magnetic islands and the sizes of the divertor footprints. This is not
sufficient to relate the sizes of the footprints to the sizes of
the islands unless the radial dependence of $\tilde{S}_n$ is known. In
this section an approximative form of this dependence at the edge 
will be given for the special case of external magnetic perturbations
imposed by coils located at the low-field side. The motivation for
this case is the use of such coils as an ELM control mechanism
\cite{evans2008} where the the coils are supposed to impact the edge
region where the ELMs originate.

We will use a simplified model of the perturbed magnetic field where
the perturbation is localized at the low field side where
the field line pitch angle $\d\varphi/\d\theta$ ($\theta$ being the
geometric poloidal angle) is assumed to be constant poloidally and
radially. This is a realistic assumption for the edge region near the
separatrix which is our region of interest.
We will
note the local pitch angle  $q_1$:
$q_1=\d\varphi/\d\theta=\mbox{const.}$ The variation of the safety
factor is assumed to be caused only by the variation of the pitch
angle in the regions where the perturbation is negligible: the
high-field side and especially the X-point. This requires the perturbation
coils to be placed sufficiently far from the X-point region.


Along a field line in the low field side region we have
\begin{equation}
\thh=o\varphi/q=oq_1 \theta/q.\label{eq:thh}
\end{equation}
It follows that the $\homoc$ function has a simple dependence on
$\theta$ in this region:
\begin{equation}
\homoc=q_1 \theta.\label{eq:homoc-geom}
\end{equation}
The $\mgeo$ Fourier component of the
perturbation w.r.t the geometric poloidal angle $\theta$ is
defined as
\begin{equation}
\label{eq:bpertgeodef}
\bpertgeo{\mgeo}{n} = \dpiint \exp(-\mbox{i}\mgeo\theta) \bpertngeo(\theta)\;\d\theta
\end{equation}
where $\bpertngeo(\theta)$ is the $n$ toroidal Fourier component of
$\delta b^1(\theta, \varphi)$ considered as a
function of $\theta$:
\begin{equation}
\bpertngeo(\theta) \equiv \bpertn(\homoc(\theta)).\label{eq:bpertngeo}
\end{equation}

We will now find the relation between the Fourier components
$\bpert{m}{n}$ and $\bpertgeo{\mgeo}{n}$. We are neglecting the
perturbation outside the region where the Eq.~(\ref{eq:homoc-geom})
holds which allows to express $\bpert{m}{n}$
[Eq.~(\ref{eq:homocnmode})] in terms of the $\theta$ coordinate:
\begin{equation}
  \bpert{m}{n} = \frac{q_1}{o2\pi q}\int_{-\pi}^{\pi}
  \exp\left[-\mbox{i}o\left(\frac{mq_1}{q}\theta\right)\right]
   \bpertngeo(\theta)\d\theta\label{eq:1}
\end{equation}
Equations (\ref{eq:1}) and (\ref{eq:bpertngeo}) finally give simple relations between
$\bpert{m}{n}$ and $\bpertgeo{\mgeo}{n}$:
\begin{equation}
  \label{eq:bperteq}
  \bpert{m}{n} = \frac{q_1}{q} \bpertgeo{mq_1/q}{n} = \frac{q_1}{q} \bpertgeo{nq_1}{n}
\end{equation}
and
$$
\bpertgeo{\mgeo}{n}  =
\frac{q}{q_1} \bpert{\mgeo q/q_1}{n}.
$$
From those it can be seen why the maxima and minima of the spectrum
$\bpert{m}{n}$ in $(m, s)$ space form ``ridges'' and ``valleys''
aligned with the $q$ profile, as can be seen e.g. for the proposed 
ITER designs in
Ref.~\onlinecite{marinanf2008} (Fig.~15c therein) and noted for DIII-D in
Ref.~\onlinecite{joseph2008} (see Fig.~1b therein). The Fourier
component w.r.t. \thh{} -- $\bpert{m}{n}$ --  is given by
Eq.~(\ref{eq:bperteq}).
Assuming that the Fourier component of
the perturbation w.r.t. $\theta$ -- $\bpertgeo{\mgeo}{n}$ --
does not change significantly between different magnetic surfaces, 
the only radial dependence is the inverse
proportionality to $q$ which is the same for all the poloidal modes. If $\bpert{m}{n}$ has the maximum
on one surface with $s=s_1$ for $m=m_{\text{max}}(s_1)$, on other
resonant surface with  $s=s_2$ it
will have also maximum for $m=m_{\text{max}}(s_2)$ equal to 
$m_{\text{max}}(s_1)q(s_2)/q(s_1)$ so maxima will be aligned with the
$q$ profile which is given in the $(m, s)$ space as the set of points
satisfying $m=nq(s)$.

Using these resuls the an approximate radial dependence of
$\bpert{m}{n}$ can be found. The radial dependence of
the geometric poloidal Fourier component $\tilde{B}^{\prime
  r}_{\mgeo}$ of the radial perturbation $B^r\equiv\delta
\vek{B}\cdot\vek{e}_r$ (with $\vek{e}_r$ being the unit vector
perpendicular to the magnetic surfaces)
is\cite{fitzpatrick:3337} $\tilde{B}^{\prime
  r}_{\mgeo}\propto r^{\mgeo-1}$. The contravariant $s$ component $B^1$
is given by $B^1=B^r\partial s/\partial r$. 
Assuming that $\partial
s/\partial r$ and $B^3$ do not depend significantly on the poloidal
angle in the area with a non-negligible perturbation, the geometric poloidal
Fourier component of the normalized contravariant perturbation is
given by
\begin{equation}
  \label{eq:bpertgeorad}
  \bpertgeo{\mgeo}{n} \propto r^{\mgeo-1} \parder{s}{r} \frac{1}{B^3}.
\end{equation}
The radial dependence of $\bpert{m}{n}$ is given by the formula
(\ref{eq:bperteq}) where Eq.~(\ref{eq:bpertgeorad}) can be used to
substitute for $\bpertgeo{mq_1/q}{n}$. 

For resonant modes we may use the Melnikov-like function instead. 
From Eqs.~(\ref{eq:homoc-geom}), (\ref{eq:melnikov-like}) and
(\ref{eq:bpertgeodef}) it follows that 
\begin{equation}
 \tilde{S}_n(s) =  \frac{q_1}{o2\pi} \bpertgeo{nq_1}{n}\label{eq:melnikov-bpertgeo}
\end{equation}
The radial dependence of $\tilde{S}_n(s)$ can be obtained from
Eqs.~(\ref{eq:bpertgeorad}) and (\ref{eq:melnikov-bpertgeo}):
\begin{equation}
 \tilde{S}_n(s) \propto  q_1 r^{nq_1-1}\parder{s}{r} \frac{1}{B^3}
 \label{eq:melnikov-rad}
\end{equation}
At a sufficiently narrow edge region the right-hand side is not
strongly radially dependent, so we may expect the values of
$\tilde{S}_n(s)$ on different resonant surface to be strongly
correlated. Note that (\ref{eq:melnikov-rad}) and this conclusion
applies also to the value on separatrix $\tilde{S}_n(s=1)$, which is
thus the limit of $\tilde{S}_n(s)$ at the resonant surfaces
approaching the separatrix, because (\ref{eq:melnikov-rad}) does not
contain discontinuous terms.

\section{Conclusion}
We derived a generalized formula for analytic estimation of width of
magnetic islands which does not rely on a simplified cylindrical
geometry, but instead takes into account toroidal toroidal geometry
and arbitrary (i.e. noncircular) cross-section of magnetic surfaces.
This makes it especially suitable for estimating the edge ergodization
in an X-point tokamak geometry, where the edge region is substantially
different from a cylindrical approximation. The formula is based on
the perturbed magnetic field and we demonstrated its equivalence to
formulae expressed in terms of the perturbed poloidal flux.
We then formulated assumptions about the form of the perturbed
magnetic field which correspond to the perturbations typically used inmo
the ongoing effort to control ELMs with magnetic perturbations on a
range of tokamaks. Namely, we suppose that the perturbation acts
mostly in a region away from the X-point, where the pitch angle of the
field lines does not have a significant radial variation in the region
of interest, which is the edge zone near the separatrix. This
assumption is valid for the coils used for ELM control experiments in
most tokamaks, as well as the proposed coils for ITER. Using this
assumption we then derived more concrete results about the
perturbation harmonics which determine the island sizes. We
demonstrated that all the resonant harmonics are correlated.
Our result expresses formally the
alignment of the maxima and minima of the perturbation spectra with
the safety factor profile, which is often observed in the calculations
of perturbation harmonics. 
We also show that the quantity which determines the
island sizes is also directly linked to the Melnikov integral and thus
determines the extent of the footprints on the divertor
plates.

Our results show that by using coils on low field side it is not
possible to create significantly diffferent resonant perturbations on
different rational surfaces. Maximizing the resonant mode on one
surface also leads to maximization of resonant modes on other
surfaces. This is advantageous if one wants to optimize the coil
system for maximum island overlap and stochastization. If one rather
wants to study the effect on perturbation on each surface separately
it might be more advantageous to choose a different position of the
coils, as it is the case for the new perturbation coils on
DIII-D. Maximizing the island overlap will also lead to maximization
of divertor footprints due to the relation between island sizes and
the Melnikov integral.

As our method is restricted to a LFS-localized perturbation, the
results do not apply to a perturbation field created inside the plasma
itself, e.g. a locked mode. In this case the relation between magnetic
islands and the divertor footprints may be much less constrained.

\begin{acknowledgments}
This work was partly supported by the European Communities under the
contracts of association between EURATOM, IPP.CR and CEA. The
views and opinions expressed herein do not
necessarily reflect those of the European Commission.
\end{acknowledgments}

\appendix

\section{Hamiltonian representation of field lines and magnetic
  islands}
\label{sec:hamilt-repr-field}
{ 
In the theory of hamiltonian dynamical systems (see
Ref.~\onlinecite{lichtlieber}), the formula for the island width is derived
using the hamiltonian description of field line dynamics, with the
poloidal flux function in the role of the hamiltonian and the toroidal
angle in the role of time (see e.g. Ref.~\onlinecite{abdullaev:042508}). The
hamiltonian is defined as
\begin{equation}
  \label{eq:ham-pot}
  H = A_\varphi = RA_{\hat{\varphi}}
\end{equation}
where $A_\varphi$ is the covariant toroidal component of the vector
potential and $A_{\hat{\varphi}} = \vek{A}\cdot\enorm{\varphi}$ is the
physical component, with $\enorm{\varphi}$ being the unit basis vector in
the toroidal direction. A convenient choice of canonical coordinates
is the action-angle representation, where the action is the toroidal
flux $\toroflux$ and the angle is $\thh$. The Hamiltonian equations
are:
\begin{eqnarray}
  \label{eq:hameq1}
  \der{\thh}{\varphi} = \parder{H}{\toroflux}\\
  \label{eq:hameq2}
  \der{\toroflux}{\varphi} = -\parder{H}{\thh}
\end{eqnarray}

In the equilibrium case $H$ is a function of poloidal position only
and is independent on the toroidal angle $\varphi$. Moreover,
$\toroflux$ and $\thh$ being action-angle variables, they are chosen
so that $H$ is only a function of $\toroflux$ and
Eq.~(\ref{eq:hameq2}) is identically zero. A nonaxisymmetric
perturbation is represented by the addition of a small term
$\epsilon{}H_1(\toroflux, \thh, \varphi)$ to the hamiltonian, which
can then be written as
\begin{equation}
  \label{eq:pertham}
  H(\toroflux, \thh, \varphi) = H_0(\toroflux) +
  \epsilon{}H_1(\toroflux, \thh, \varphi).
\end{equation}
$H_0$ is the equilibrium part, which can be identified with the
unperturbed poloidal flux $\psi$. The perturbed part
$\epsilon{}H_1(\toroflux, \thh, \varphi)$ corresponds to a
perturbation $\delta A_\varphi(\toroflux, \thh, \varphi)$ of
$A_\varphi$. The equilibrium part has the property
\begin{equation}
  \label{eq:derflux}
  \der{H_0}{\toroflux}=\der{\psi}{\toroflux} = \frac{1}{q}
\end{equation}
which reduces Eq.~(\ref{eq:hameq1}) to the form (\ref{eq:thline}). (We
assume that the perturbation term $\epsilon{}\der{H_1(\toroflux, \thh,
  \varphi)}{\toroflux}$ is negligible in comparison with the
equilibrium term $1/q$ and can be neglected.) The
equation~(\ref{eq:sline}) can be derived from (\ref{eq:hameq2}) by
expressing the perturbed field $\delta b^1$ using the perturbed
potential $\delta A_\varphi$. This expression is 
\begin{equation}
\label{eq:pert-pot}
\delta b^1 = -\der{s}{\psi}\frac{1}{q}\parder{\delta A_\varphi}{\thh}.
\end{equation}
The derivative
$\der{s}{\varphi}$ can be expressed as
\begin{equation}
  \label{eq:ders-derpsi}
  \der{s}{\varphi} =
  \der{s}{\psi}\der{\psi}{\toroflux}\der{\toroflux}{\varphi}=
  \der{s}{\psi}\frac{1}{q}\der{\toroflux}{\varphi}.
\end{equation}
From Eqs.~(\ref{eq:pert-pot}) and (\ref{eq:ders-derpsi}) it follows
that Eq.~(\ref{eq:sline}) is equivalent to (\ref{eq:hameq2}).

It is useful to decompose the perturbed potential in Fourier modes,
analogously to the decomposition (\ref{eq:bmnsum}) of $\delta b^1$:
\begin{eqnarray}
  \label{eq:hammnsum}
  \delta A_\varphi&=&\epsilon{}H_1=\epsilon\sum_{m,n=-\infty}^{\infty}
  {\Hpert{m}{n}\exp{[\mbox{i}(m\thh-n\varphi)]}}\\
  &=& \epsilon\sum_{m,n}
  {\Hpertreal{m}{n}\cos(m\thh-n\varphi+\chi_{mn})}
\end{eqnarray}
From (\ref{eq:bmnsum}), (\ref{eq:hammnsum}) and (\ref{eq:pert-pot}) we
obtain the relation between $\bpert{m}{n}$ and $\Hpert{m}{n}$:
\begin{eqnarray}
  \label{eq:bpermn-Hmn}
  \bpert{m}{n} &=& - \der{s}{\psi}\frac{1}{q}
  \mbox{i}m\epsilon{}\Hpert{m}{n}\\
  \label{eq:bpertreal-Hpertreal}
  \bpertreal{m}{n}&=&\der{s}{\psi}\frac{1}{q}m\epsilon{}\Hpertreal{m}{n}.
\end{eqnarray}

The half-width of islands measured in terms of the action variable
(toroidal flux $\toroflux$) is\cite{lichtlieber}:
\begin{equation}
  \label{eq:linhalfwidth_ham}
  \delta_\toroflux = 2q\sqrt{\frac{\epsilon{}\Hpertreal{m}{n}}{\der{q}{\toroflux}}}
\end{equation}
In a linear approximation, the half-width in terms of $s$ is
related to $\delta_\toroflux$ by the relation
$\delta_s=\der{s}{\toroflux}\delta_\toroflux$. Moreover,
$\der{q}{\toroflux}=\der{s}{\toroflux}q^\prime$ and $\der{s}{\toroflux}=\der{s}{\psi}\frac{1}{q}$, so using
Eq.~(\ref{eq:bpertreal-Hpertreal}) we see that the expressions
(\ref{eq:linhalfwidth_ham}) and (\ref{eq:linhalfwidth}) are
equivalent.
}

\bibliography{/home/pavel/bibliografie/my.bib}
\end{document}